\documentclass[twocolumn,showpacs,preprintnumbers,amsmath,amssymb,ctexart]{revtex4}
\usepackage{tipa}
\usepackage{graphicx}
\usepackage{graphics}
\usepackage{subfigure}
\usepackage{dcolumn}
\usepackage{slashbox}
\usepackage{bm}
\usepackage{amsmath}
\usepackage{amssymb}
\begin{document}

\linespread{1.5}
\title{Hadronic decays of the highly excited $2D$ $D_s$ resonances}
\author{Jing Ge$^1$\footnote{xiaofeige91@126.com}, Dan-Dan Ye$^{1,2}$ and Ailin Zhang$^1$
\footnote{Corresponding author:zhangal@staff.shu.edu.cn}}
\affiliation{$^1$ Department of Physics, Shanghai University, Shanghai 200444, China\\
$^2$ College of Mathematics, Physics and Information Engineering, Jiaxing University
,Jiaxing 314001, China}

\begin{abstract}
Hadronic decays of the highly excited $2D$ $D_s$ resonances have been studied in the $^3P_0$ model. Widths of all possible hadronic decay channels of the $2D$ $D_s$ have been computed. $D^*_{s1}(2700)$, $D^*_{s1}(2860)$, $D^*_{s3}(2860)$, $D(2600)$ and $D(2750)$ can be produced from hadronic decays of the $2D$ $D_s$, and relevant hadronic decay widths have been particularly paid attention to. The hadronic decay widths of $2D$ $D_s$ to $D(2600)$ or $D(2750)$ may be large, and the numerical results are different in different assignments of $D(2600)$ and $D(2750)$. The hadronic decay widths of $2D$ $D_s$ to $D^*_{s1}(2860)$, $D^*_{s3}(2860)$ or $D^*_{s1}(2700)$ are very small, and different in different assignments of $D^*_{s1}(2700)$.
\end{abstract}
\pacs{13.25.Ft\\
Keywords: Hadronic decay, $^3P_0$ model}
\maketitle
\section{Introduction}
\label{intro}
\par The properties of highly excited heavy-light meson states have been studied for a long time, the S-wave and P-wave $D$ and $D_s$ resonances are believed established. In recent years, more and more heavy-light resonances such as the higher excited $D_s$ have been observed. However, some candidates of these highly excited heavy-light resonances have not been definitely pinned down.

\par $D^*_{s1}(2700)$ was first observed by $Belle$~\cite{belle} in $B^+ \rightarrow \bar{D}^0D_{s1} \rightarrow \bar{D}^0D^0K^+$. It was also observed by $BaBar$ in $D^*K$ channel~\cite{babar}. This state is included in PDG~\cite{pdg} with $M=2709\pm 4 ~\rm{MeV}$, $J^P=1^-$ and $~\Gamma=117\pm 13 ~\rm{MeV}$.

\par $D^*_{sJ}(2860)$ was first reported by $BaBar$~\cite{babar} in $D_{s1}(2860) \rightarrow D^0K^+,D^+K^0$ with a mass $M=2856.6\pm1.5~\rm(stat)\pm5.0~\rm(syst){MeV}$ and a width $~\Gamma=48\pm7 ~\rm(stat)\pm10~\rm(syst){MeV}$. It was observed once again in the $D^*K$ channel~\cite{babar}. This state is included in PDG~\cite{pdg} with $M=2863.2^{+4.0}_{-2.6}{MeV}$, $~\Gamma=58\pm11~\rm{MeV}$ and unknown $J^P$.
Recently, LHCb collaboration reported that the resonance $m(\bar D^0K^-\approx 2.86)$ GeV contain both spin-1 and spin-3 components~\cite{lhcb,lhcb2}. These two states, $D^*_{s1}(2860)^-$ and $D^*_{s3}(2860)^-$, are suggested to be the $J^P=1^-$ and $J^P=3^-$ members of the $1D$ family. In addition, two new charmed states, $D(2600)$ and $D(2750)$, were observed by $BaBar$ Collaboration~\cite{bb,pdg}.

\par $D^*_{s1}(2700)$ was supposed the first radially excited S-wave states $D_s(2^3S_1)$~\cite{FEColse,rizzi,chen,chen2}, the orbitally excited $D_s(1^3D_1)$~\cite{Liu,Godfrey} or their mixture~\cite{chen,zhang,FEColse}. Similarly, $D^*_{sJ}(2860)^\pm$ was once suggested as the $J^P=0^+$~\cite{rupp,FEColse}, $J^P=3^-$~\cite{chen,3-,Godfrey,zhao} excited $D_s$ or the orthogonal partner of $D^*_{s1}(2700)$\cite{zhang,FEColse,Ma,Godfrey}. The $0^+$ possibility is subsequently excluded by the observation of $D_{s1}(2860) \rightarrow D^*K$ channel. Recent experiment suggests that there are in fact two $D^*_{s1}(2860)^-$ and $D^*_{s3}(2860)^-$ close to $D^*_{sJ}(2860)^\pm$ ~\cite{lhcb,lhcb2}, where $D^*_{s1}(2860)^-$ and $D^*_{s3}(2860)^-$ are suggested to be the $J^P=1^-$ and $J^P=3^-$ members of the $1D$ family~\cite{lhcb,lhcb2,Godfrey}.

$D(2600)$ is observed and suggested to be the first radially excited S-wave states $D(2^3S_1)$ through an analyse of their masses and helicity-angle distribution, while $D(2750)$ is observed and suggested the orbitally excited $1D$ state~\cite{bb}. In Ref.~\cite{liu3,zhong}, $D(2600)$ is suggested as an admixture of $2^3S_1$ and $1^3D_1$ with $J^P=1^-$, and $D(2750)$ is interpreted as an orthogonal partner of $D(2600)$ or $1^3D_3$. In Ref.~\cite{chen2}, $D(2600)$ was interpreted as a pure $2^3S_1$ state from its hadronic decays in the heavy quark symmetry theory. In Ref.~\cite{colangelo}, an analysis of the assignment of $D(2600)$ to the first radial excitation of $D^*$ has been done in detail in an effective Lagrangian approach.

There are different interpretations to these resonances. Obviously, the nature of these resonances have not been understood clearly. In literatures, the arrangements of these resonances are mainly based on the study of their $J^P$ quantum numbers, masses and strong decay modes.

 It is well known that the study of productions of these resonances is also an important way to understand them. $D^*_{s1}(2700)$, $D^*_{s1}(2860)$, $D^*_{s3}(2860)$, $D(2600)$ and $D(2750)$ can be produced from the strong decays of highly excited resonances. It will be interesting to study the hadronic production of $D^*_{s1}(2700)$, $D^*_{s1}(2860)$, $D^*_{s3}(2860)$, $D(2600)$ and $D(2750)$ from higher excited resonances. In fact, some highly excited $D_s$ resonances have been observed by BaBar, LHCb et al., more and more highly excited $D_s$ resonances are expected to be observed by these Collaborations. For kinematical reason, these resonances can be produced from hadronic decays of $2D$ $D_s$. Unfortunately, the strong decays of the highly excited $2D$ $D_s$ resonances have seldom been studied before. In this paper, the hadronic decays of these $2D$ $D_s$ resonances will be studied in the $^3P_0$ model.

\par The paper is organized as follows. In Sec.II, we give a brief review of the $^3P_0$ model and possible decay modes of the $2D$ resonances. In Sec. III, we present the formula and numerical results of the hadronic decay of the $2D$ $D_s$ resonances, and the decays with $D^*_{s1}(2700)$, $D^*_{s1}(2860)$, $D^*_{s3}(2860)$, $D(2600)$ or $D(2750)$ involved in the final states are particularly paid attention to. Finally, the conclusions and discussions are given in Sec. IV.

\section{$^3P_0$ MODEL AND POSSIBLE DECAY MODES OF THE $2D$ $D_s$ RESONANCES}
$^3P_0$ model is popularly known as a quark-pair creation (QPC) model, which has been extensively applied to the calculation of the OZI-allowed strong decay of meson $A$ to meson $B$ and $C$. The model was first proposed by Micu~\cite{micu1969}, and then developed by Yaouanc et al~\cite{yaouanc1,yaouanc2,ac}. The decay process is shown in Fig. 1~\cite{liu2,zhang3}, where a pair of quarks $q_3\bar{q_4}$ with $J^{PC}=0^{++}$ are created from the vacuum and regroup with the $q_1\bar{q_2}$ within the initial meson $A$ into two outgoing mesons $B$ and $C$.

\par
\begin{figure}[h]
\begin{center}
\includegraphics[height=2.8cm,angle=0,width=6cm]{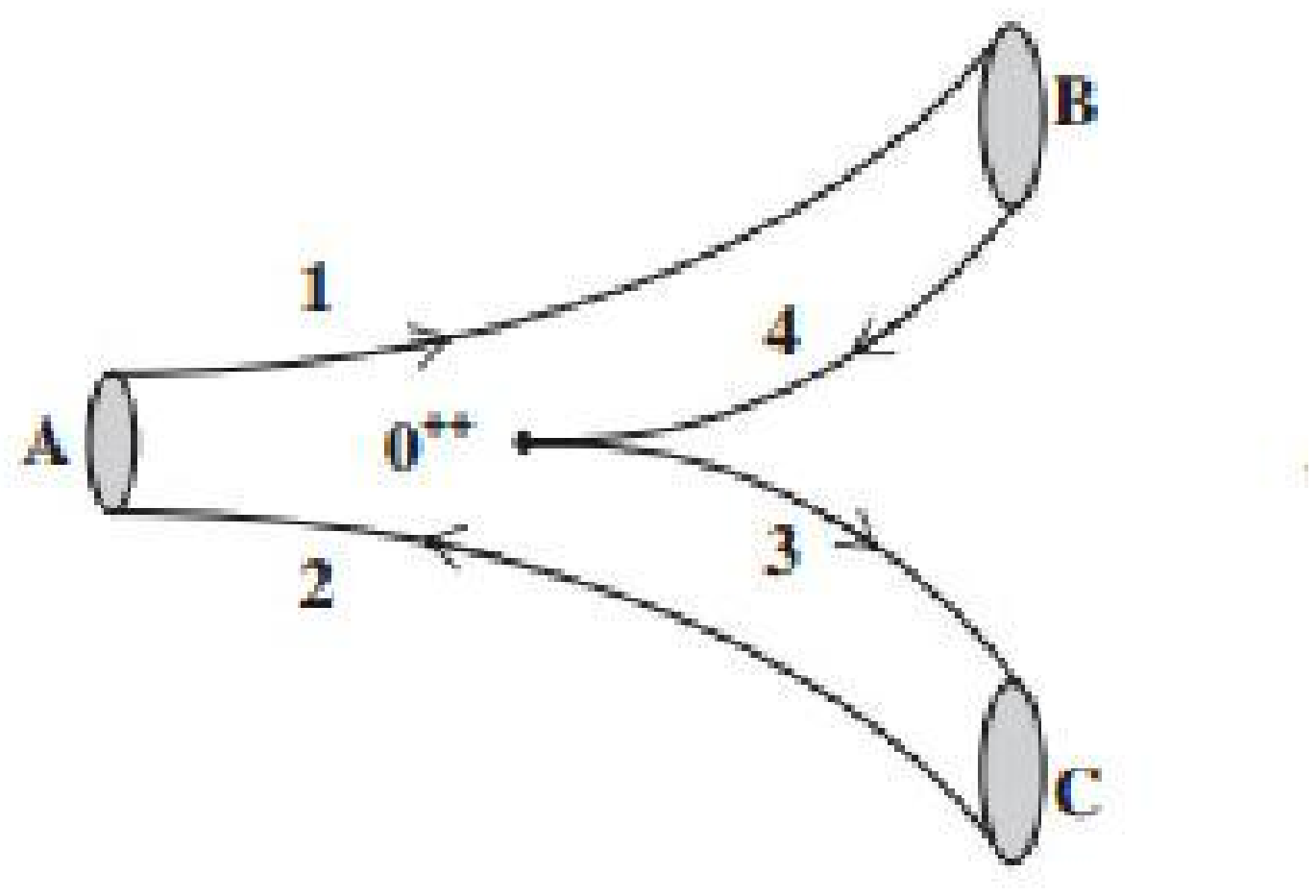}
\caption{Decay process of $A\to B+C$ in the $^3P_0$ model.}
\end{center}
\end{figure}

In $^3P_0$ model, the hadronic decay width of $A\to BC$ is
\begin{eqnarray}
\Gamma  = \pi ^2 \frac{|\vec{K}|}{M_A^2}\sum_{JL} |{\mathcal{M}^{JL}}|^2
\end{eqnarray}
where the momentum of the daughter meson in the initial meson A's center of mass frame is
\begin{eqnarray}
 |\vec{K}|=\frac{{\sqrt {[m_A^2-(m_B-m_C )^2][m_A^2-(m_B+m_C)^2]}}}{{2m_A }}
\end{eqnarray}
and $M^{JL}$ is the partial wave amplitude of $A \rightarrow BC$. In terms of the Jacob-Wick formula~\cite{JW}, the partial wave amplitude can be obtained from the helicity amplitude $\mathcal{M}^{M_{J_A } M_{J_B } M_{J_C }}$

\begin{flalign}
\mathcal{M}^{JL} (A \to BC) &= \frac{{\sqrt {2L + 1} }}{{2J_A  + 1}} \nonumber \\
&\times\sum_{M_{J_B } ,M_{J_C } } \langle {L0JM_{J_A } } |{J_A M_{J_A } }\rangle  \nonumber \\
&\times\langle {J_B M_{J_B } J_C M_{J_C } } |J, {JM_{J_A } } \rangle \nonumber \\
 &\times \mathcal{M}^{M_{J_A } M_{J_B } M_{J_C } } (\vec{K})
\end{flalign}
with $\vec{J}=\vec{J_B}+\vec{J_C}$, $\vec{J_A}=\vec{J_B}+\vec{J_C}+\vec{L}$ and $M_{J_A}=M_{J_B}+M_{J_C}$. In this equation, the helicity amplitude
\begin{flalign}
 &\mathcal{M}^{M_{J_A } M_{J_B } M_{J_C }}\nonumber \\
 &=\sqrt {8E_A E_B E_C } \gamma \sum_{\mbox{\tiny$\begin{array}{c}
M_{L_A } ,M_{S_A } ,\\
M_{L_B } ,M_{S_B } ,\\
M_{L_C } ,M_{S_C } ,m\end{array}$}}  \langle {L_A M_{L_A } S_A M_{S_A } }| {J_A M_{J_A } }\rangle \nonumber \\
 &\times\langle L_B M_{L_B } S_B M_{S_B }|J_B M_{J_B } \rangle \langle L_C M_{L_C } S_C M_{S_C }|J_C M_{J_C }\rangle\nonumber \\
 & \times \langle {1m;1 - m}|{00} \rangle\langle \chi _{S_B M_{S_B }}^{13} \chi _{S_C M_{S_C } }^{24}|\chi _{S_A M_{S_A } }^{12} \chi _{1 - m}^{34}\rangle \nonumber \\
&\times\langle\varphi _B^{13} \varphi _C^{24}|\varphi _A^{12}\varphi _0^{34} \rangle I_{M_{L_B } ,M_{L_C } }^{M_{L_A },m} (\vec{K})
\end{flalign}
while the spatial integral $I_{M_{L_B },M_{L_C}}^{M_{L_A },m}(\vec{K})$ is
\begin{flalign}
I_{M_{L_B } ,M_{L_C } }^{M_{L_A } ,m} (\vec{K})&= \int d \vec{k}_1 d \vec{k}_2 d \vec{k}_3 d \vec{k}_4 \nonumber \\
&\times\delta ^3 (\vec{k}_1 + \vec{k}_2-\vec{p}_A)\delta ^3 (\vec{k}_3+ \vec{k}_4)\nonumber \\
&\times \delta ^3 (\vec{p}_B- \vec{k}_1- \vec{k}_3 )\delta ^3 (\vec{p}_C- \vec{k}_2 -\vec{k}_4) \nonumber \\
& \times\Psi _{n_B L_B M_{L_B } }^* (\vec{k}_1 ,\vec{k}_3)\Psi _{n_cL_C  M_{L_c}}^* (\vec{k}_2 ,\vec{k}_4) \nonumber \\
& \times \Psi _{n_A L_A M_{LA}} (k_1 ,k_2 )Y _{1m}\left(\frac{\vec{k_3}-\vec{k}_4}{2}\right).
\end{flalign}

The details of the indices, matrix elements and other indications are given in Ref.~\cite{zhang3}

\par With these formula in hand, we go ahead with our calculation. In the calculation, the simple harmonic oscillator(SHO) wave function is employed to represent the meson wave function. The meson flavor functions follow the convention in Ref.~\cite{flavor}: $D^0=c\bar{u}$, $D^+=-c\bar{d}$, $D_s^+=-c\bar{s}$, $K^+=-u\bar{s}$, $K^-=s\bar{u}$, $\phi=-s\bar{s}$, $\eta=(u\bar{u}-d\bar{d})/2-s\bar{s}/\sqrt{2}$ and $\eta\prime=(u\bar{u}-d\bar{d})/2+s\bar{s}/\sqrt{2}$.

For the parameters involved in $^3P_0$ model, the light nonstrange quark pair creation strength $\gamma$ and the strange quark pair creation strength $\gamma_{s\bar{s}}$ are correlated by $\gamma_{s\bar{s}}\approx\gamma/\sqrt{3}$~\cite{yaouanc2} with $\gamma=7.85$~\cite{Lx3040}. The constituent quarks masses are taken to be $m_c=1.43$ GeV, $m_u=m_d=0.45$ GeV and $m_s=0.55$ GeV~\cite{Ma}. The resonance masses and the effective scale parameters $\beta$ for different resonances used in our calculation are listed in Table. 1~\cite{pdg,Ma} and Table. 2~\cite{Ma}, respectively. There is not a $2D$ $D_s$ observed, and the masses of these resonances are unknown. In our calculation, theoretical predicted masses of the $1^-$ $2^3D_1$ $D_s$ (3383 MeV) and the $3^-$ $2^3D_3$ $D_s$ (3469 MeV)~\cite{RT} are employed, respectively. For the $2^-$ $D_s$ resonance, $2^3D_2$ may mix with $2^1D_2$, which may result in a complicated mixing. Only when the detail of the mixing is clear, can we give the hadronic decay widths of each $2^-$ resonances. In this paper, we give only the results of pure $2^3D_2$ and $2^1D_2$. As an approximation, the average mass ($3429.5$ MeV) of the two predicted $2^-$ $D_s$ in Ref.~\cite{RT} is taken as the mass input of $2^3D_2$ and $2^1D_2$.
\begin{table}[b]
\caption{Meson masses used in our calculation (MeV).}
\begin{tabular}{p{0.4cm}p{1.5cm}*{2}{p{1.8cm}}p{1.3cm}}
   \hline\hline
   & States&Mass& States&Mass\\
   \hline
    &$K^{\pm\rm}$ & $493.677$ & $K_1(1270)^{0(\pm\rm)}$   & $1272$    \\
   &$K^0$        & $497.614$ & $K_1(1400)^{0(\pm\rm)}$   & $1403$    \\
   &$K^{*\pm\rm}$& $891.66$  & $K_2^*(1430)^0$           & $1432.4$  \\
   &$K^{*0}$     & $896$     & $K_2^*(1430)^{\pm\rm}$    & $1425.6$  \\
   &$\eta$       & $547.853$ & $K_0^*(1430)^{0(\pm\rm)}$ & $1425$    \\
   &$\eta\prime$ & $957.78$  &$K^*(1410)^{0(\pm\rm)}$   &$1414$     \\
   &$\phi$       & $1019.455$& $D_1(2430)^{0(\pm\rm)}$   & $2427$  \\
   &$D^{\pm\rm}$  & $1869.2$  &  $D_1(2420)^{0}$           & $2422.3$ \\
   &$D^0$          & $1864.84$  & $D_1(2420)^{\pm\rm}$      & $2423.4$ \\
   &$D^{*\pm\rm}$  & $2010.27$ & $D_0(2400)^{0}$           & $2308$   \\
   &$D^{*0}$       & $2006.97$  & $D_0(2400)^{\pm\rm}$      & $2403$   \\
   &$D_s$          & $1968.49$ & $D_2(2460)^{\pm\rm}$      & $2460.1$ \\
   &$D_S^*$        & $2112.3$    & $D_2(2460)^{0}$           & $2461.1$\\
   &$D_{s0}(2317)$ & $2317.8$    & $D(2550)^{0(\pm\rm)}$     & $2539.4$ \\
   &$D_{s1}(2460)$ & $2459.6$  & $D(2600)^0$               & $2608.7$  \\
   &$D_{s1}(2536)$ & $2535.35$ & $D(2600)^{\pm\rm}$        & $2621.3$\\
   &$D_{s2}(2573)$ & $2571.9$  & $D(2750)^0$               & $2763.3$    \\
   &$D^*_{s1}(2700)$ & $2709$  & $D(2750)^{\pm\rm}$        & $2769.7$   \\
   &$D^*_{s1}(2860)$ & $2859$   &$D^*_{s3}(2860)$          &$2860.5$\\
  
   \hline\hline
\end{tabular}
\label{table1}
\end{table}

\begin{table}[t]
\caption{Different $\beta$ values for the S-wave, P-wave and D-wave resonances in MeV.}
\begin{tabular}{p{0.3cm}*{6}{p{1.2cm}}}
   \hline\hline
   &$n^{2S+1}L_J$ & $u\bar{u}$ & $u\bar{s}$ & $s\bar{s}$ & $c\bar{u}$ & $c\bar{s}$\\
   \hline
   &$1^1S_0$           &470&466&470&453&484    \\
   &$2^1S_0$           &294&301&310&325&343    \\
   &$1^3S_1$           &308&322&338&379&406  \\
   &$2^3S_1$           &258&267&279&306&324  \\
   &$1^3P_J$           &280&290&302&328&348    \\
   &$2^3P_J$           &247&255&265&287&303   \\
   &$1^1P_1$           &284&294&306&332&352  \\
   &$2^1P_1$           &250&259&269&290&306  \\
   &$1^3D_J$           &261&270&281&304&321   \\
   &$2^3D_J$           &238&246&255&275&290  \\
   &$1^1D_2$           &261&270&281&304&321  \\
   &$2^1D_2$           &238&246&255&275&290 \\
   \hline\hline
\end{tabular}
\label{table2}
\end{table}
Possible kinematically allowed decay modes of these four $2D$ $D_s$ resonances are presented in Table. 3 and Table. 4.
\begin{table*}[]
\caption{OZI-allowed hadronic decay modes of $D_s(2^3D_1)$ and $D_s(2^3D_3)$. The masses of $D_s(2^3D_1)$ and $D_s(2^3D_3)$ are $3383$ MeV and $3469$ MeV, respectively~\cite{RT}. }
\begin{tabular}{p{0cm}p{2cm}p{6cm}p{2cm}p{4.5cm}}
   \hline\hline
   &&$D_s(2^3D_1)$& &$D_s(2^3D_3)$\\
   \hline
   &Mode &Channels &Mode &Channels\\
   \hline
   & $0^{+}+1^{-}$  & $D(2400)^0K^{*+},D(2400)^+K^{*0},$       & $0^{+}+1^{-}$  & $D(2400)^0K^{*+},D(2400)^+K^{*0},$\\
    & $           $  & $D_s(2317)\phi$                          & $           $  & $D_s(2317)\phi$ \\
    & $0^{-}+1^{+}$  & $D^0K_1(1270)^+,D^+K_1(1270)^0$          & $0^{-}+1^{+}$  & $D^0K_1(1270)^+,D^+K_1(1270)^0,$\\
    & $           $  & $D^0K_1(1400)^+,D^+K_1(1400)^0$          & $           $  & $D^0K_1(1400)^+,D^+K_1(1400)^0$\\
    & $1^{+}+0^{-}$  & $D(2430)^0K^{+},D(2430)^+K^{0},$         & $1^{-}+0^{+}$  & $D^{*0}K_0^*(1430)^+,D^{*+}K_0^*(1430)^0$\\
    & $           $  & $D(2420)^0K^{+},D(2420)^+K^{0},$         & $1^{+}+0^{-}$  &$D(2430)^0K^{+},D(2430)^+K^{0},$\\
    & $           $  & $D_{s1}(2460)\eta,D_{s1}(2536)\eta$       & $           $  &$D(2420)^0K^{+},D(2420)^+K^{0},$\\
    & $1^{+}+1^{-}$  & $D(2430)^0K^{*+},D(2430)^+K^{*0}$         & $           $  & $D_{s1}(2460)\eta,D_{s1}(2460)\eta\prime,$ \\
    & $           $  &$D(2420)^0K^{*+},D(2420)^+K^{*0}$          & $           $  &$D_{s1}(2536)\eta$ \\
    & $           $  & $                                $        & $1^{+}+1^{-}$  &$D(2430)^0K^{*+},D(2430)^+K^{*0},$\\
    & $1^{-}+1^{+}$  & $D^{*0}K_1(1270)^+,D^{*+}K_1(1270)^0$      & $           $  &$D(2420)^0K^{*+},D(2420)^+K^{*0}$\\
    & $2^{+}+1^{-}$  &$D_2(2460)^0K^{*+},D_2(2460)^+K^{*0}$       & $1^{-}+1^{+}$  &$D^{*0}K_1(1270)^+,D^{*+}K_1(1270)^0,$\\
    & $0^{-}+0^{-}$  &$D^0K^+,D^+K^0,D_s\eta$                     & $           $  & $D^{*0}K_1(1400)^+,D^{*+}K_1(1400)^0$\\
    & $           $  & $D(2550)^0K^+,D(2550)^+K^0$                &  $2^{+}+1^{-}$  &$D_2(2460)^0K^{*+},D_2(2460)^+K^{*0}$\\
    & $1^{-}+0^{-}$  &$D^{*0}K^+,D^{*+}K^0,$                        & $1^{-}+2^{+}$  &$D^{*0}K_2(1430)^+,D^{*+}K_2(1430)^0$\\
    & $           $  &$D_s^*\eta,D_s^*\eta\prime,D_s(2^3S_1)\eta,$  & $0^{-}+0^{-}$  &$D^0K^+,D_s\eta,D(2550)^0K^+,$\\
    & $           $   &$D(2^3S_1)^0K^+,D(2^3S_1)^+K^0,$           & $           $  & $D^+K^0,D(2550)^+K^0$ \\
    & $           $   &$D(1^3D_1)^0K^+,D(1^3D_1)^+K^0$            & $1^{-}+0^{-}$  &$D^{*0}K^+,D^{*+}K^0,D_s^*\eta,D_s^*\eta\prime,$\\
    & $0^{-}+1^{-}$  &$D^{0}K^{*+},D^{+}K^{*0},D_s\phi$           & $           $  & $D_s(2^3S_1)\eta,D_s(1^3D_1)\eta,$ \\
    &                &$D^{0}K^*(1410)^+,D^{+}K^*(1410)^0 $        & $           $  &$D(2^3S_1)^0K^+,D(2^3S_1)^+K^0,$\\
    & $1^{-}+1^{-}$  &$D^{*0}K^{*+},D^{*+}K^{*0},D_s^*\phi$       & $           $  & $D(1^3D_1)^0K^+,D(1^3D_1)^+K^0$ \\
    & $2^{+}+0^{-}$  &$D_2(2460)^0K^+,D_2(2460)^+K^0,$            & $0^{-}+1^{-}$  &$D^{0}K^{*+},D^{+}K^{*0},D_s\phi$\\
    & $           $  &$D_{s2}(2573)\eta$                          &                &$D^{0}K^*(1410)^+,D^{+}K^*(1410)^0 $\\
    & $0^{-}+2^{+}$  &$D^0K_2^{*}(1430)^+,D^+K_2^{*}(1430)^0$     & $1^{-}+1^{-}$  &$D^{*0}K^{*+},D^{*+}K^{*0},D_s^*\phi$\\
    &                 &                                           &                &$D^{*0}K^*(1410)^+,D^{*+}K^*(1410)^0$\\
    & $           $  &$                                 $         & $2^{+}+0^{-}$  &$D_2(2460)^0K^+,D_2(2460)^+K^0,$\\
    & $           $  &$                                 $         & $           $  &$D_{s2}(2573)\eta$ \\
    & $           $  &$                                 $         & $0^{-}+2^{+}$  &$D^0K_2^{*}(1430)^+,D^+K_2^{*}(1430)^0$\\
    & $           $  &$                                 $         & $3^{-}+0^{-}$  &$D(1^3D_3)^0K^+,D(1^3D_3)^+K^0,$\\
    & $           $  &$                                 $         & $           $  & $D_s(1^3D_3)\eta$ \\
\hline\hline
\end{tabular}
\label{table3}
\end{table*}

\begin{table*}[]
\caption{OZI-allowed hadronic decay modes of $D_s(2^1D_2)$ and $D_s(2^3D_2)$. The mass of $D_s(2^1D_2)$ and $D_s(2^3D_2)$ is $3429.5$ MeV ~\cite{RT}. }
\begin{tabular}{p{0cm}p{2cm}p{6cm}p{2cm}p{4.5cm}}
   \hline\hline
   &&$D_s(2^1D_2)$& &$D_s(2^3D_2)$\\
   \hline
   &Mode &Channels &Mode &Channels\\
   \hline
   & $0^{+}+1^{-}$  & $D(2400)^0K^{*+},D(2400)^+K^{*0},$           & $0^{+}+1^{-}$  & $D(2400)^0K^{*+},D(2400)^+K^{*0},$\\
   & $           $  & $D_s(2317)\phi$                              & $           $  & $D_s(2317)\phi$ \\
   & $0^{-}+1^{+}$  & $D^0K_1(1270)^+,D^+K_1(1270)^0$              & $0^{-}+1^{+}$  & $D^0K_1(1270)^+,D^+K_1(1270)^0,$\\
   & $           $  & $D^0K_1(1400)^+,D^+K_1(1400)^0$              & $           $  & $D^0K_1(1400)^+,D^+K_1(1400)^0$\\
   & $1^{+}+0^{-}$  & $D(2430)^0K^{+},D(2430)^+K^{0},$             & $1^{+}+0^{-}$  &$D(2430)^0K^{+},D(2430)^+K^{0},$\\
   & $           $  & $D(2420)^0K^{+},D(2420)^+K^{0},$             & $           $  &$D(2420)^0K^{+},D(2420)^+K^{0},$\\
   & $           $  & $D_{s1}(2460)\eta,D_{s1}(2536)\eta$          & $           $  & $D_{s1}(2460)\eta,D_{s1}(2460)\eta\prime,$ \\
   & $1^{+}+1^{-}$  & $D(2430)^0K^{*+},D(2430)^+K^{*0}$            & $           $  &$D_{s1}(2536)\eta$ \\
   & $           $  &$D(2420)^0K^{*+},D(2420)^+K^{*0}$             & $1^{+}+1^{-}$  &$D(2430)^0K^{*+},D(2430)^+K^{*0},$\\
   & $           $  & $                                $           & $           $  &$D(2420)^0K^{*+},D(2420)^+K^{*0}$\\
   & $1^{-}+1^{+}$  & $D^{*0}K_1(1270)^+,D^{*+}K_1(1270)^0$        & $1^{-}+1^{+}$  &$D^{*0}K_1(1270)^+,D^{*+}K_1(1270)^0,$\\
   & $           $  & $D^{*0}K_1(1400)^+,D^{*+}K_1(1400)^0$        & $           $  & $D^{*0}K_1(1400)^+,D^{*+}K_1(1400)^0$\\
   & $2^{+}+1^{-}$  &$D_2(2460)^0K^{*+},D_2(2460)^+K^{*0}$         & $2^{+}+1^{-}$  &$D_2(2460)^0K^{*+},D_2(2460)^+K^{*0}$\\
   & $0^{+}+0^{-}$  &$D(2400)^0K^+,D_s(2317)\eta,$                 & $0^{+}+0^{-}$  &$D(2400)^0K^+,D_s(2317)\eta,$\\
   & $           $  & $D(2400)^+K^0,D_s(2317)\eta\prime  $         & $           $  & $D(2400)^+K^0,D_s(2317)\eta\prime$ \\
   & $0^{-}+0^{+}$  &$D^{0}K^*_0(1430)^+,D^{+}K^*_0(1430)^0$       & $0^{-}+0^{+}$  &$D^{0}K^*_0(1430)^+,D^{+}K^*_0(1430)^0$\\
   & $1^{-}+0^{-}$  &$D^{*0}K^+,D^{*+}K^0,$                        & $1^{-}+0^{-}$  &$D^{*0}K^+,D^{*+}K^0,D_s^*\eta,D_s^*\eta\prime,$\\
   & $           $  &$D_s^*\eta,D_s^*\eta\prime,D_s(2^3S_1)\eta,$  & $           $  & $D_s(2^3S_1)\eta,D_s(1^3D_1)\eta,$ \\
   & $           $  &$D(2^3S_1)^0K^+,D(2^3S_1)^+K^0,$              & $           $  &$D(2^3S_1)^0K^+,D(2^3S_1)^+K^0,$\\
   & $           $  &$D(1^3D_1)^0K^+,D(1^3D_1)^+K^0$               & $           $  & $D(1^3D_1)^0K^+,D(1^3D_1)^+K^0$ \\
   & $0^{-}+1^{-}$  &$D^{0}K^{*+},D^{+}K^{*0},D_s\phi$             & $0^{-}+1^{-}$  &$D^{0}K^{*+},D^{+}K^{*0},D_s\phi$\\
   &                &$D^{0}K^*(1410)^+,D^{+}K^*(1410)^0 $          &                 &$D^{0}K^*(1410)^+,D^{+}K^*(1410)^0 $\\
   & $1^{-}+1^{-}$  &$D^{*0}K^{*+},D^{*+}K^{*0},D_s^*\phi$         & $1^{-}+1^{-}$  &$D^{*0}K^{*+},D^{*+}K^{*0},D_s^*\phi$\\
   & $2^{+}+0^{-}$  &$D_2(2460)^0K^+,D_2(2460)^+K^0,$              &                 &$D^{*0}K^*(1410)^+,D^{*+}K^*(1410)^0$\\
   & $           $  &$D_{s2}(2573)\eta$                            & $2^{+}+0^{-}$  &$D_2(2460)^0K^+,D_2(2460)^+K^0,$\\
   & $0^{-}+2^{+}$  &$D^0K_2^{*}(1430)^+,D^+K_2^{*}(1430)^0$       & $           $  &$D_{s2}(2573)\eta$ \\
   & $3^{-}+0^{-}$  &$D(1^3D_3)^0K^+,D(1^3D_3)^+K^0,$              & $0^{-}+2^{+}$  &$D^0K_2^{*}(1430)^+,D^+K_2^{*}(1430)^0$\\
   & $           $  & $D_s(1^3D_3)\eta$                            & $3^{-}+0^{-}$  &$D(1^3D_3)^0K^+,D(1^3D_3)^+K^0,$\\
   & $           $  &$                                 $           & $           $  & $D_s(1^3D_3)\eta$ \\

\hline\hline
\end{tabular}
\label{table4}
\end{table*}

\section{HADRONIC DECAYS OF $2D$ $D_s$ RESONANCES}

\par Possible hadronic decay modes and their numerical decay widths of $D_s(2^3D_1)$ and $D_s(2^3D_3)$ except for final states including $D^*_{s1}(2700)$, $D*_{s1}(2860)$, $D*_{s3}(2860)$, $D(2600)$ or $D(2750)$ are shown in Table. 5. Similar results of the strong decays of $D_s(2^3D_2)$ and $D_s(2^1D_2)$ are shown in Table. 6.

In our calculation, $D_1(2430)$ and $D_{s1}(2460)$ are assigned as the $1^+(j^P=\frac{1}{2}^+)$ $D$ and $D_s$, respectively. $D_1(2420)$ and $D_{s1}(2536)$ are assigned as the excited $1^+(j^P=\frac{3}{2}^+)$ $D$ and $D_s$. Through the relation between the $j^P$ eigenstates and the $^{2S+1}L_J$ eigenstates, these two $1^+$ resonances are regarded as a mixture of $1^1P_1$ and $1^3P_1$ resonances
\begin{eqnarray*}
\left(
\begin{array}{c}
|1^+,j^p=\frac{1}{2}^+\rangle\\
  \\
|1^+,j^p=\frac{3}{2}^+\rangle
\end{array}
\right)=\left(\begin{array}{c}
cos\theta  \quad  sin\theta\\
\displaystyle
-sin\theta  \quad cos\theta   \\
\end{array}
\right)\left(
\begin{array}{c}
|1^1P_1\rangle\\
    \displaystyle
|1^3P_1 \rangle
\end{array}
\right)
\end{eqnarray*}
where the mixing angle $\theta=-tan^{-1}\sqrt{2}=-54.7^\circ$~\cite{RT,Lx3040}.

\begin{table*}[t]
\caption{Hadronic decay widths of $D_s(2^3D_1)$ and $D_s(2^3D_3)$ in MeV.}
\begin{tabular}{p{0cm}p{2.2cm}p{1.2cm}p{2.2cm}p{2cm}p{2.2cm}p{1.2cm}p{2.2cm}p{1.2cm}}
   \hline\hline
   &&$D_s(2^3D_1)$& &&&$D_s(2^3D_3)$\\
   \hline
   & Channels& Width& Channels& Width& Channels& Width& Channels& Width\\
   \hline
   &$D(2400)^0K^{*+}$    & $13.53$          &$D^+K^0$             & $0.24$    &$D(2400)^0K^{*+}$    & $6.10$         &$D^{*+}K_2(1430)^0$   & $53.04$\\
   & $D(2400)^+K^{*0}$   & $7.28$           & $D_s\eta$           & $0.05$     &$D(2400)^+K^{*0}$    & $7.48$         & $D^0K^+$             & $0.00$ \\
   &$D_s(2317)\phi$      & $2.83$           &$D_s\eta\prime$      & $1.03$     &$D_s(2317)\phi$      & $2.19$         & $D^+K^0$            & $0.00$   \\
   &$D^0K_1(1270)^+$     & $14.88$          & $D(2550)^0K^+$      & $12.34$    &$D^0K_1(1270)^+$     & $81.61$         &$D_s\eta$           & $0.00$  \\
   &$D^+K_1(1270)^0$     & $14.58$          & $D(2550)^+K^0$      & $12.55$    &$D^+K_1(1270)^0$     & $75.06$        &$D_s\eta\prime$      & $0.16$  \\
   &$D^0K_1(1400)^+$     & $15.29$          &$D^{*0}K^+$          & $3.91$     &$D^0K_1(1400)^+$     & $33.16$         &$D(2550)^0K^+$      & $0.06$\\
   &$D^+K_1(1400)^0$     & $13.32$          & $D^{*+}K^0$         & $4.08$     &$D^+K_1(1400)^0$     & $36.30$         & $D(2550)^+K^0$     & $0.06$ \\
   & $D(2430)^0K^{+}$    & $20.88$          &$D_s^*\eta$          & $1.60$     &$D^{*0}K_0^*(1430)^+$   & $1.03$       &$D^{*0}K^+$          & $0.00$\\
    &$D(2430)^+K^{0}$     & $21.18$         & $D_s^*\eta\prime$   & $0.48$     &$D^{*+}K_0^*(1430)^0$   & $0.84$       & $D^{*+}K^0$         & $0.00$ \\
    & $D(2420)^0K^{+}$    & $13.91$         & $D^{0}K^{*+}$       & $6.31$     &$D(2430)^0K^{+}$     & $14.00$         & $D_s^*\eta$         & $0.00$\\
    &$D(2420)^+K^{0}$     & $14.04$         &$D^{+}K^{*0}$        & $6.55$     &$D(2430)^+K^{0}$     & $13.98$         & $D_s^*\eta\prime$   & $0.95$ \\
    & $D_{s1}(2460)\eta$  & $4.78$          & $D_s\phi$           & $2.20$     &$D(2420)^0K^{+}$     & $1.93$          &$D^{0}K^{*+}$        & $0.01$ \\
    &$D_{s1}(2536)\eta$   & $2.33$          &$D^{0}K^*(1410)^+$   & $20.22$     &$D(2420)^+K^{0}$     & $1.97$          & $D^{+}K^{*0}$       & $0.01$ \\
    & $D(2430)^0K^{*+}$   & $11.41$         &$D^{+}K^*(1410)^0$   & $20.19$     &$D_{s1}(2460)\eta$   & $1.34$          &$D_s\phi$            & $0.00$ \\
    &$D(2430)^+K^{*0}$    &$11.26$          & $D^{*0}K^{*+}$      & $9.82$     &$D_{s1}(2460)\eta\prime$ & $0.01$      &$D^{0}K^*(1410)^+$   &$38.98$ \\
    &$D(2420)^0K^{*+}$    & $11.33$         & $D^{*+}K^{*0}$      & $9.67$     &$D_{s1}(2536)\eta$     & $0.59$        &$D^{+}K^*(1410)^0 $  &$37.16$ \\
    &$D(2420)^+K^{*0}$    & $10.01$         &$D_s^*\phi$          & $2.76$      & $D(2430)^0K^{*+}$    & $15.27$       &$D^{*0}K^*(1410)^+$  &$68.97$\\
    &$D^{*0}K_1(1270)^+$  & $22.02$         &$D_2(2460)^0K^+$     & $0.05$     &$D(2430)^+K^{*0}$     & $14.64$        &$D^{*+}K^*(1410)^0$  &$66.56$\\
    &$D^{*+}K_1(1270)^0$  & $21.12$         &$D_2(2460)^+K^0$     & $0.04$     &$D(2420)^0K^{*+}$     & $19.17$        &$D^{*0}K^{*+}$       & $2.91$ \\
    &$D_2(2460)^0K^{*+}$  & $10.20$         &$D_{s2}(2573)\eta$   & $0.21$     &$D(2420)^+K^{*0}$     & $18.90$        & $D^{*+}K^{*0}$      & $2.93$  \\
    &$D_2(2460)^+K^{*0}$  & $10.26$         &$D^0K_2^{*}(1430)^+$ & $0.67$     &$D^{*0}K_1(1270)^+$   & $36.47$        &$D_s^*\phi$          & $0.37$\\
   & $D^0K^+$            & $0.23$           &$D^+K_2^{*}(1430)^0$ & $0.68$     &$D^{*+}K_1(1270)^0$   & $36.51$        & $D_2(2460)^0K^+$    & $0.14$  \\
    &                     &                 &                     &            &$D^{*0}K_1(1400)^+$   & $4.31$         &$D_2(2460)^+K^0$     & $0.14$ \\
    &                     &                 &                     &            &$D^{*+}K_1(1400)^0$   & $3.87$         & $D_{s2}(2573)\eta$  & $0.01$  \\
    &                     &                 &                     &            &$D_2(2460)^0K^{*+}$   & $25.44$        &$D^0K_2^{*}(1430)^+$  & $0.61$ \\
    &                     &                 &                     &            &$D_2(2460)^+K^{*0}$   & $25.54$        & $D^+K_2^{*}(1430)^0$  & $0.62$  \\
    &                     &                 &                     &            &$ D^{*0}K_2(1430)^+$  & $61.94$        &                       &       \\
   \hline\hline
\end{tabular}
\label{table5}
\end{table*}
\begin{table*}[t]
\caption{Hadronic decay widths of $D_s(2^1D_2)$ and $D_s(2^3D_2)$ in MeV.}
\begin{tabular}{p{0cm}p{2.2cm}p{1.2cm}p{2.2cm}p{2cm}p{2.2cm}p{1.2cm}p{2.2cm}p{1.2cm}}
   \hline\hline
   &&$D_s(2^1D_2)$& &&&$D_s(2^3D_2)$\\
   \hline
   & Channels& Width& Channels& Width& Channels& Width& Channels& Width\\
   \hline
   &$D(2400)^0K^{*+}$    & $0.12$          &$D(2400)^+K^0$       & $6.76$      &$D(2400)^0K^{*+}$    & $5.55$         &$D^{+}K^*(1430)^0$   & $0.02$\\
   & $D(2400)^+K^{*0}$   & $0.23$          & $D_s(2317)\eta$     & $0.87$      &$D(2400)^+K^{*0}$    & $7.68$         & $D(2400)^0K^+$       & $3.58$ \\
   &$D_s(2317)\phi$      & $0.04$          &$D_s(2317)\eta\prime$& $0.61$      &$D_s(2317)\phi$      & $1.80$         & $D(2400)^+K^0$       & $1.84$   \\
   &$D^0K_1(1270)^+$     & $0.25$          &$D^{0}K^*_0(1430)^+$ & $10.87$     &$D^0K_1(1270)^+$     & $30.32$         &$D_s(2317)\eta$      & $0.54$  \\
   &$D^+K_1(1270)^0$     & $0.20$          &$D^{+}K^*_0(1430)^0$ & $10.60$     &$D^+K_1(1270)^0$     & $30.43$         &$D_s(2317)\eta\prime$ & $0.30$  \\
   &$D^0K_1(1400)^+$     & $0.16$          &$D^{*0}K^+$          & $50.30$     &$D^0K_1(1400)^+$     & $17.50$         &$D^{*0}K^+$          & $35.95$\\
   &$D^+K_1(1400)^0$     & $0.17$          & $D^{*+}K^0$         & $49.46$     &$D^+K_1(1400)^0$     & $17.23$         & $D^{*+}K^0$         & $35.53$ \\
   & $D(2430)^0K^{+}$    & $1.50$          &$D_s^*\eta$          & $21.32$     &$D(2430)^0K^{+}$     & $1.21$          & $D_s^*\eta$         & $9.53$\\
    &$D(2430)^+K^{0}$     & $1.50$         & $D_s^*\eta\prime$   & $8.59$      &$D(2430)^+K^{0}$     & $1.12$          & $D_s^*\eta\prime$   & $6.09$ \\
    & $D(2420)^0K^{+}$    & $0.80$         & $D^{0}K^{*+}$       & $23.57$     &$D(2420)^0K^{+}$     & $14.26$          &$D^{0}K^{*+}$        & $20.41$ \\
    &$D(2420)^+K^{0}$     & $0.78$         &$D^{+}K^{*0}$        & $22.58$     &$D(2420)^+K^{0}$     & $14.12$          & $D^{+}K^{*0}$       & $20.03$ \\
    & $D_{s1}(2460)\eta$  & $0.16$          &$D^{0}K^*(1410)^+$  & $45.32$     &$D_{s1}(2460)\eta$   & $0.00$          &$D_s\phi$            & $2.32$ \\
    &$D_{s1}(2460)\eta\prime$&$0.00$        &$D^{+}K^*(1410)^0$  & $44.17$     &$D_{s1}(2460)\eta\prime$ & $1.40$      &$D^{0}K^*(1410)^+$   &$47.75$ \\
    &$D_{s1}(2536)\eta$   & $0.03$          & $D_s\phi$          & $1.60$      &$D_{s1}(2536)\eta$     & $0.80$        &$D^{+}K^*(1410)^0 $  &$47.16$ \\
    & $D(2430)^0K^{*+}$   & $21.33$         &$D^{*0}K^*(1410)^+$ & $5.83$      & $D(2430)^0K^{*+}$    & $11.58$        &$D^{*0}K^*(1410)^+$  &$2.92$\\
    &$D(2430)^+K^{*0}$    &$20.13$         &$D^{*+}K^*(1410)^0$  & $2.93$      &$D(2430)^+K^{*0}$     & $10.73$        &$D^{*+}K^*(1410)^0$  &$1.46$\\
    &$D(2420)^0K^{*+}$    & $22.74$          & $D^{*0}K^{*+}$    & $25.46$     &$D(2420)^0K^{*+}$     & $19.74$        &$D^{*0}K^{*+}$       & $15.67$ \\
    &$D(2420)^+K^{*0}$    & $21.62$         & $D^{*+}K^{*0}$     & $25.12$     &$D(2420)^+K^{*0}$     & $18.75$        & $D^{*+}K^{*0}$      & $14.91$  \\
    &$D^{*0}K_1(1270)^+$  & $49.69$         &$D_s^*\phi$         & $3.92$      &$D^{*0}K_1(1270)^+$   & $34.58$        &$D_s^*\phi$          & $2.67$\\
    &$D^{*+}K_1(1270)^0$  & $49.74$        &$D_2(2460)^0K^+$     & $19.98$     &$D^{*+}K_1(1270)^0$   & $34.13$        & $D_2(2460)^0K^+$    & $17.32$  \\
     &$D^{*0}K_1(1400)^+$  & $29.63$        &$D_2(2460)^+K^0$    & $19.98$     &$D^{*0}K_1(1400)^+$   & $43.44$         &$D_2(2460)^+K^0$     & $17.45$ \\
     &$D^{*+}K_1(1400)^0$  & $30.20$       &$D_{s2}(2573)\eta$   & $2.68$      &$D^{*+}K_1(1400)^0$   & $44.63$         & $D_{s2}(2573)\eta$  & $3.66$  \\
    &$D_2(2460)^0K^{*+}$  & $14.08$         &$D^0K_2^{*}(1430)^+$& $23.23$     &$D_2(2460)^0K^{*+}$   & $6.93$          &$D^0K_2^{*}(1430)^+$  & $15.22$ \\
    &$D_2(2460)^+K^{*0}$  & $13.84$        &$D^+K_2^{*}(1430)^0$ & $19.93$     &$D_2(2460)^+K^{*0}$    & $7.04$         & $D^+K_2^{*}(1430)^0$  & $11.49$  \\
    &$D(2400)^0K^+$       & $6.39$         &                     &             &$ D^{0}K^*(1430)^+$   & $0.03$          &                       &       \\

   \hline\hline
\end{tabular}
\label{table6}
\end{table*}

For a particular $J$, the larger the angular momentum ($L$) between the two final states, the smaller the corresponding $M^{JL}$. However, for a particular decay channel, both $J$ and $L$ could vary, and there is not an one-to-one relation between the decay width and the $M^{JL}$. From the numerical results in Table. 5, the dominant decay modes of $D_s(2^3D_1)$ are $D(2430)K$, $D^*K_1(1270)$ and $DK^*(1410)$ et al, while the dominant decay modes of $D_s(2^3D_3)$ are $DK_1(1270)$, $D^*K^*(1410)$ and $D^*K_2(1430)$ et al. From the numerical results in Table. 6, the dominant decay modes of $D_s(2^1D_2)$ are $D^*K$, $D^*K_1(1270)$ and $DK^*(1410)$ et al, while the dominant decay modes of $D_s(2^3D_2)$ are $DK^*(1410)$, $D^*K_1(1400)$ and $D^*K$ et al. In forthcoming experiments, the $2D$ $D_s$ resonances are expected to be observed in these dominant hadronic decay channels.

\par Since there are different assignments to $D^*_{s1}(2700)$, $D(2600)$ and $D(2750)$, their hadronic productions (together with the production of $D^*_{s1}(2860)$ and $D^*_{s3}(2860)$) from $2D$ $D_s$ resonances are studied independently in the following subsection.

\subsection{$D^*_{s1}(2700)$, $D^*_{s1}(2860)$ and $D^*_{s3}(2860)$}

\par For kinematical reason and conservation of some quantum numbers in hadronic decay, $D^*_{s1}(2700)$ can only be produced through $2D~D_s\to D^*_{s1}(2700)\eta$. The numerical results are presented in Table. 7. The first column in the table indicates three possible assignments of $D^*_{s1}(2700)$, where the mixture possibility is from Ref.~\cite{Godfrey} with a mixing angle $\theta=88^\circ$. The mixing of $D^*_{s1}(2700)$ has also been studied in other references~\cite{chen,Ma,yuan}. In the table, all the decay widths are very small though they are different in different assignments of $D^*_{s1}(2700)$.

\begin{table*}[t]
\caption{Hadronic decay widths (in MeV) of $2D$ $D_s\to D^*_{s1}(2700)\eta$ in different assignments of $D^*_{s1}(2700)$, where the mixture is from Ref.~\cite{Godfrey}.}

\begin{tabular}{p{0cm}p{3.2cm}p{3.2cm}p{3.2cm}p{3.2cm}p{3.2cm}p{3.2cm}}
   \hline\hline
    &$State$        &$2^3D_1\rightarrow D^*_{s1}(2700)\eta$ &$2^3D_3\rightarrow D^*_{s1}(2700)\eta$ & $2^1D_2\rightarrow D^*_{s1}(2700)\eta$&$2^3D_2\rightarrow D^*_{s1}(2700)\eta$\\
    \hline
    &$D^*_{s1}(2700)(2^3S_1)$ &$2.48$                           &$1.43$                         &$3.98$                          &$6.26$\\
    &$D^*_{s1}(2700)(1^3D_1)$ &$0.78$                           &$0.03$                         &$8.39$                          &$3.19$\\
    &$D^*_{s1}(2700)(\rm{mixture})$ & $0.87$                    &$0.02$                         &$8.08$                          &$3.50$\\
\hline\hline
\end{tabular}
\label{table7}
\end{table*}

\par Once $D^*_{s1}(2860)^-$ and $D^*_{s3}(2860)^-$ are assigned as the $J^P=1^-$ and $J^P=3^-$ members of the $1D$ family, the hadronic decay widths of $2D~D_s\to D_{sJ}(2860)\eta$ can also be calculated. However, predicted masses of $2D$ $^3D_1$, $^3D_2$ and $^1D_2$ $D_s$~\cite{RT} are close to the threshold of $D^*_{s1}(2860)/D^*_{s3}(2860)\eta$, their hadronic decays may be complicated. Therefore, we give only the results of the hadronic decay channels $D_s (2^3D_3)\rightarrow D^*_{s1}(2860)/D^*_{s3}(2860)\eta$. The decay widths for $D_s (2^3D_3)\rightarrow D^*_{s1}(2860)\eta$ MeV and $D_s (2^3D_3)\rightarrow D^*_{s3}(2860)\eta$ are $~0$ and $0.37$ MeV, respectively.

\subsection{$D(2600)$ AND $D(2750)$}

$D(2600)$ is possibly a $2^3S_1$, or a $1^3D_1$ $D$, or an orthogonal partner of the mixtures of $2^3S_1$ and $1^3D_1$ with $J^P=1^-$~\cite{chen,Ma,liu2}
\begin{eqnarray*}
|(SD)_1\rangle_L=cos\theta|2^3S_1\rangle-sin\theta|1^3D_1\rangle\\
|(SD)_1\rangle_R=sin\theta|2^3S_1\rangle+cos\theta|1^3D_1\rangle
\end{eqnarray*}

The hadronic decay widths of $2D$ $D_s\to D(2600)K$ in different assignments of $D(2600)$ are given in Table. 8. Obviously, the decay widths have large difference in different assignments of $D(2600)$, therefore the final decay widths depend heavily on the mixing angle. In the table, results corresponding to two different mixing angle are given, where the mixing angle $0.9\leq\theta\leq1.5$~\cite{Ma} and $0.36\leq\theta\leq0.40$~\cite{liu2}.

\begin{table*}[t]
\caption{Hadronic decay widths (in MeV) of $2D$ $D_s\to D(2600)K$ in different assignments of $D(2600)$.}

\begin{tabular}{p{0cm}p{3.2cm}p{3.2cm}p{3.2cm}p{3.2cm}p{3.2cm}p{3.2cm}}
   \hline\hline
    &$Assignment\backslash Mode$        &$2^3D_1\rightarrow D(2600)^+K^0$ &$2^3D_3\rightarrow D(2600)^+K^0$ & $2^1D_2\rightarrow D(2600)^+K^0$ & $2^3D_2\rightarrow D(2600)^+K^0$\\
    \hline
    &$D(2600)^+(2^3S_1)$ &$9.85$                           &$22.27$                        &$28.80$                          &$26.35$\\
    &$D(2600)^+(1^3D_1)$ &$36.20$                          &$0.87$                         &$138.02$                         &$52.26$\\
    &$D(2600)^+(\rm{mixture}$\cite{Ma}$)$ &$37.99-45.33$     &$0.00-4.86$                    &$90.81-136.75$                   &$40.85-51.93$\\
    &$D(2600)^+(\rm{mixture}$\cite{liu2}$)$ &$25.53-27.14$   &$15.87-16.63$                  &$39.22-41.67$                    &$28.68-29.25$\\
    \hline
    &$Assignment\backslash Mode$        &$2^3D_1\rightarrow D(2600)^0K^+$ &$2^3D_1\rightarrow D(2600)^0K^+$ & $2^3D_3\rightarrow D(2600)^0K^+$ & $2^3D_3\rightarrow D(2600)^0K^+$\\
    \hline
    &$D(2600)^0(2^3S_1)$ &8.89                           &$22.96$                        &$29.10$                          &$25.76$\\
    &$D(2600)^0(1^3D_1)$  &42.35                         &$0.82$                         &$145.99$                         &$56.35$\\
    &$D(2600)^0(\rm{mixture}$\cite{Ma}$)$ &$44.92-51.24$    &$0.00-5.15$                   &$99.58-145.22$                   &$38.37-53.97$\\
    &$D(2600)^0(\rm{mixture}$\cite{liu2}$)$ &$26.03-27.88$  &$16.50-17.27$                 &$43.07-45.91$                    &$25.45-25.92$\\
\hline\hline
\end{tabular}
\label{table7}
\end{table*}
$D(2750)$ may be a $1^3D_1$ or a $1^3D_3$ $D$, numerical results of relevant channels are given in Table. 9, where the "-" indicates an impossible channel. As indicated in Ref.~\cite{chen2}, there are possibly two $D$ resonances close to $2750$ MeV, which requires careful study of this energy region. Therefore, other possible assignments of $D(2750)$ are not studied here.

\begin{table*}[t]
\caption{Hadronic decay widths (in MeV) of $2D$ $D_s\to D(2750)K$ in different assignments of $D(2750)$.}

\begin{tabular}{p{0cm}p{3.2cm}p{3.2cm}p{3.2cm}p{3.2cm}p{3.2cm}p{3.2cm}}
   \hline\hline
    &$Assignment\backslash Mode$        &$2^3D_1\rightarrow D(2750)^+K^0$ &$2^3D_3\rightarrow D(2750)^+K^0$ & $2^1D_2\rightarrow D(2750)^+K^0$ & $2^3D_2\rightarrow D(2750)^+K^0$\\
    \hline
    &$D(2750)^+(1^3D_3)$ &$-$                           &$6.02$                         &$45.59$                          &$52.32$\\
    &$D(2750)^+(1^3D_1)$ &$1.91$                           &$0.44$                         &$48.27$                          &$20.64$\\
    \hline
    &$Assignment\backslash Mode$        &$2^3D_1\rightarrow D(2750)^0K^+$ &$2^3D_3\rightarrow D(2750)^0K^+$ & $2^3D_3\rightarrow D(2750)^0K^+$ & $2^3D_3\rightarrow D(2750)^0K^+$\\
    \hline
    &$D(2750)^0(1^3D_3)$ &$-$                           &$5.81$                         &$49.98$                          &$56.97$\\
    &$D(2750)^0(1^3D_1)$  &2.58                       &$0.49$                         &$53.22$                          &$22.28$\\
\hline\hline
\end{tabular}
\label{table7}
\end{table*}

\section{CONCLUSIONS AND DISCUSSIONS}
\par In this work, the hadronic decay properties of the highly excited $2D$ $D_s$ resonances are studied in $^3P_0$ model. The hadronic decay widths of all possible OZI-allowed channels of the highly excited $2D$ $D_s$ resonances have been calculated. The dominant decay modes of $D_s(2^3D_1)$ are $D(2430)K$, $D^*K_1(1270)$ and $DK^*(1410)$ et al, while the dominant decay modes of $D_s(2^3D_3)$ are $DK_1(1270)$, $D^*K^*(1410)$ and $D^*K_2(1430)$ et al. The dominant decay modes of $D_s(2^1D_2)$ are $D^*K$, $D^*K_1(1270)$ and $DK^*(1410)$ et al, while the dominant decay modes of $D_s(2^3D_2)$ are $DK^*(1410)$, $D^*K_1(1400)$ and $D^*K$ et al. The $2D$ $D_s$ resonances are suggested to be observed in these dominant hadronic decay channels in forthcoming experiments.

\par $D^*_{s1}(2700)$, $D^*_{s1}(2860)$ and $D^*_{s3}(2860)$ can be produced from the hadronic decays of the highly excited $2D$ $D_s$ resonances. Possible hadronic decay channels are $2D$ $D_s\to D^*_{s1}(2700)\eta$ and $2D$ $D_s\to D^*_{s1}(2860)/D^*_{s3}(2860)\eta$, respectively. In every possible assignments of $D^*_{s1}(2700)$, all the hadronic decay widths are very small. The hadronic decay widths of $D_s(2^3D_3)\to D^*_{s1}(2860)\eta$ and $D_s(2^3D_3)\to D^*_{s3}(2860)\eta$ are also very small. It is not suitable to classify these resonances according to their hadronic production from the $2D$ $D_s$. The threshold of $D^*_{s1}(2860)/D^*_{s3}(2860)\eta$ are close to the theoretical predicted masses of $2D$ $^3D_1$, $^3D_2$ and $^1D_2$, and relevant hadronic decays may be complicated.

\par Hadronic decay widths of $2D$ $D_s\to D(2600)K$ and $2D$ $D_s\to D(2750)K$ in different assignments of $D(2600)$ and $D(2750)$ have also been calculated. The hadronic decay widths may be large, and the numerical results are different in different assignments of $D(2600)$ and $D(2750)$. If the $2D$ $D_s$ resonances are observed in forthcoming experiments, the measure of these hadronic decay widths will help us to understand $D(2600)$ and $D(2750)$.

\par In our paper, the uncertainties of the input parameters and the model have not been studied. The detail of possible mixing of some resonances has neither been explored. More theoretical study of these highly excited resonances are required. Of course, the most important thing is to expect more highly excited $D_s$ resonances observed in forthcoming experiments.

\begin{acknowledgments}
This work is supported by National Natural Science Foundation of China under the grants: 11075102 and 11475111. It is also supported by the Innovation Program of Shanghai Municipal Education Commission under the grant No. 13ZZ066.
\end{acknowledgments}

\end{document}